\documentclass[aps,pre,10pt,onecolumn,noshowpacs,amsfonts,amssymb,amsmath,longbibliography,superscriptaddress,notitlepage]{revtex4-1}
\usepackage[utf8]{inputenc}
\usepackage{eqnarray}
\usepackage{bm}
\usepackage[hidelinks]{hyperref}
\usepackage{graphicx}
\usepackage[dvipsnames]{xcolor}
\usepackage{subcaption}
\usepackage{array}

\begin{document}


\title{Within-host infection dynamics with master equations and the method of moments: A case study of human papillomavirus in the epithelium}

\author{Mariah C. \surname{Boudreau}}
\affiliation{Vermont Complex Systems Center, University of Vermont, Burlington VT, USA}
\affiliation{Department of Mathematics \& Statistics, University of Vermont, Burlington VT, USA}
\author{Jamie A. \surname{Cohen}}
\affiliation{Global Health Division, Institute
for Disease Modeling, Bill \& Melinda Gates Foundation, Seattle, WA, USA}
\author{Laurent \surname{H\'{e}bert-Dufresne}}
\affiliation{Vermont Complex Systems Center, University of Vermont, Burlington VT, USA}
\affiliation{Department of Mathematics \& Statistics, University of Vermont, Burlington VT, USA}
\affiliation{Department of Computer Science, University of Vermont, Burlington VT, USA}

\date{\today}

\begin{abstract} 

Master equations provide researchers with the ability to track the distribution over possible states of a system. From these equations, we can summarize the temporal dynamics through a method of moments. These distributions and their moments capture the stochastic nature of a system, which is essential to study infectious diseases. In this paper, we define the states of the system to be the number of infected cells of a given type in the epithelium, the hollow organ tissue in the human body. Epithelium found in the cervix provides a location for viral infections to live and persist, such as human papillomavirus (HPV). HPV is a highly transmissible disease which most commonly affects biological females and has the potential to progress into cervical cancer. By defining a master equation model which tracks the infected cell layer dynamics, information on disease extinction, progression, and viral output can be derived from the method of moments. From this methodology and the outcomes we glean from it, we aim to inform differing states of HPV infected cells, and assess the effects of structural information for each outcome. 

\end{abstract}

\maketitle

\section{\label{sec:intro}Introduction}
Population models for the spread of disease are a major area of study which emphasize the macroscopic effect of a disease: the number of cases, complications, hospitalizations, or mortality within a population as a whole. As we refine these models, we also aim to capture the nuances of underlying within-host disease dynamics, which are often hidden under parameters and assumptions of population models. The intricacies of within-host interactions have been explored by many, as reviewed by Speranza \cite{speranza2023understanding}. Sequencing cells, identifying infection, tracking replication, and understanding the spatial aspect of cellular infection play into the complexities that can inform heterogeneous dynamics in a population \cite{speranza2023understanding}. From literature reviews of within-host biological dynamics, we attempt to understand the mechanisms of a disease. The focus of this paper is on human papillomavirus (HPV), where the literature admits gaps in knowledge. Gravitt showcases HPV knowledge limitations in latency, clearance and incidence \cite{gravitt2011known}. These limitations can lead to a single parameter that assumes homogeneity within the dynamics or calibrating the model parameters with data. This paper takes a different approach. This work aims to use master equations and method of moments to illuminate the within-host cell dynamics of HPV that pose uncertainty in population models. \cite{epstein2008model, ryser2017mechanistic}. 

Modeling healthy and unhealthy cells in an epithelium is not a novel idea. Ordinary differential equations are used in many different ways to model within-host dynamics like epidermis dynamics \cite{ciupe2017host}. Murall \emph{et al.} focus on three systems of ordinary differential equations, two of which model skin systems for unvaccinated and vaccinated hosts exposed to HPV. The final system is a compartmental model for the transmission dynamics between individuals \cite{murall2015could}. Both Sierra-Rojas \emph{et al.} and Asih \emph{et al.} define ordinary differential equations for the populations of differing layers of the epithelium \cite{sierra2022modeling, asih2016dynamics}. Another differential equation model for general epidermis turnover was developed by Ohno \emph{et al.} \cite{ohno2021computational}. Lastly, partial differential equations enter the space when Sari \emph{et al.} consider how time and age affect the progression of HPV toward cancer \cite{sari2022mathematical}. While the mechanistic aspect of an epithelial infection is included in these models, they do not encapsulate the stochasticity that comes with infections.

All the aforementioned models are defined as mean-field models, which track the average dynamics of the system. Their structure makes it easy to define changes in average states of the system, however, the distribution of possible states around the average are missed. Mean-field models do not address the stochasticity that comes with infections. In the context of cell divisions, their random nature can cause extinction events, resulting in transient infections. On the other hand, stochastic reinfection and multiple infection events can result in persistent infections. To factor in these types of infection events, models other than ordinary or partial differential equations have been explored.

Stochasticity not only integrates varying infection events, but provides perspective on the gaps in HPV knowledge \cite{gravitt2011known, ryser2017mechanistic}. Branching processes are one stochastic method, which Ryser \emph{et al.} and Beneteau \emph{et al.} each use to address the randomness of cell division dynamics for HPV clearance \cite{ryser2015hpv, beneteau2021episome}. While modelers use branching processes to define the probability of random events, master equations strike the balance of tracking all aspects of a system with stochastic dynamics. Since probability distributions inherently provide stochasticity to a model, master equations track these distributions of all states in the system are tracked over time. This method accounts for the probability transfer between states, which is detailed in Sec.~\ref{sec:me}. One example of an infection-specific master equation approach is detailed by Vaughan \emph{et al.}, who define target, HIV-infected, and virion cells, and the dynamics between them \cite{vaughan2012within}. In contrast, Clayton \emph{et al.} give an in-depth model using master equations to tracking a cell through its division process. The stochastic divisions of the base layer cells, either asymmetric and symmetric, maintain homeostasis in an arbitrary epithelium \cite{clayton2007single}. This paper follows Clayton \emph{et al.}'s model for the epithelial division process, however, their work focuses on the homeostasis achieved in epidermis tissue and general cell clone-size distributions \cite{clayton2007single}. Our results focus on how moments of distributions computationally simplifies the issue of solving a large master equation system. Avoiding this computational expense allows for additional cell types to be added in the system, meaning structural comparisons can occur. These moments are defined through the method of moments, discussed in detail in Sec. ~\ref{sec:mom}, and allow for determining extinction, persistence, and viral output events for different structured systems.

Now, HPV is a sexually transmitted infection that is highly transmissible, leading to either transient or persistent infections. Transmission occurs through direct skin contact, causing infections to affect either the skin or mucosal epithelium. There are two categories of HPV infections: high and low-risk genotypes.  Each type comes with different symptoms, for example, low-risk genotype infections lead to skin warts. High-risk genotype infections most commonly lead epithelial lesions, which can progress into cancer. These lesions and cancer are usually found in the cervix, but cancers of the vulva, vagina, penis, anus, mouth and throat are also possible. The Centers for Disease Control report that 99\% of cervical precancers detect high-risk genotypes, and that one of these genotypes specifically cause about half of cervical cancers around the world \cite{meites2020human}. 

\begin{figure}
    \centering
    \includegraphics[width = 0.85\linewidth]{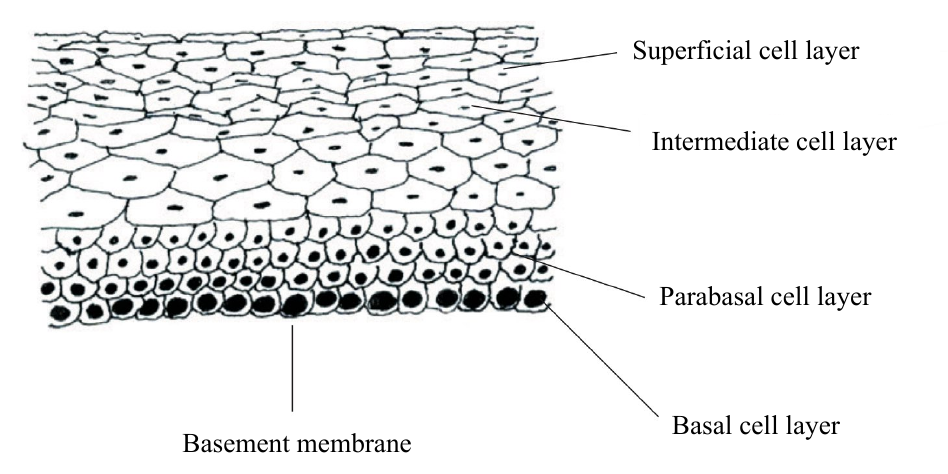}
    \caption{\textbf{Cervical epithelium:} Shown is the side view of the cervical epithelium, with its distinct cell layers. For the models presented in this paper, the basement membrane will not be included. Cells either divide or differentiate, meaning they either duplicate or transition to the subsequent layer, to sustain each layers depth \cite{clayton2007single}\footnote{Cell image reprinted from Colposcopy and Treatment of Cervical Precancer,  IARC technical publication No. 45, Walter Prendiville and Rengaswamy Sankaranarayanan, Anatomy of the uterine cervix and the transformation zone, Page 14, 2017 (2017) \cite{prendiville2017colposcopy}.}.}
    \label{fig:skin_diagram}
\end{figure}

Figure ~\ref{fig:skin_diagram} depicts a cervical epithelium composed of \textit{basal}, \textit{parabasal}, \textit{intermediate} and \textit{superficial} cell layers. The basal cells mature through each layer to the superficial layer \cite{murall2019epithelial, prendiville2017colposcopy}. In a healthy epithelium, basal and parabasal cells have the ability to divide, however, parabasal cells can also differentiate into intermediate cells. Once a cell is in the intermediate cell layer, it can only transition up to the superficial cell layer. Eventually, a superficial cell will shed and no longer be a part of the epithelium. An individual becomes infected with HPV when the virus infects the basal cell layer. The infection then propagates through cell divisions and transitions, but will only clear when there are no more infected basal cells  \cite{sierra2022modeling}.

With master equations providing the structure of differential equations and incorporating stochasticity into the system, we model HPV infection progression through epithelial tissue. We show the details of master equations and the method of moments in Sections ~\ref{sec:me}, and ~\ref{sec:mom}. From this mathematical framework, we capture the mechanistic essence of extinction, persistence, and viral load in Sections ~\ref{sec:extinction_math}, and ~\ref{sec:viral_load}. In this paper, we simulate and derive analytical measures for two different systems in Sections ~\ref{sec:sims}, ~\ref{sec:three}, and ~\ref{sec:five}. Finally, Sec. ~\ref{sec:results} will validate the method of moments with simulations, along with showcasing the structural effect on each outcome. 

\section{\label{sec:methods} Methods}

\subsection{\label{sec:assumptions} Assumptions}

In order to model this system mechanistically, we assert the following assumptions for HPV infections: First, our focus is only on specific divisions and cell processes which are described in Sections ~\ref{sec:three} and ~\ref{sec:five}. Second, our model assumes there is only one infection occurring at a time, defined through initial conditions and lack of reinfection events, even though that might not be true for a real-world scenario and could be easily relaxed in our equations. Third, it is assumed that we are infecting the system with a high-risk genotype infection, to simulate infected cells that accumulate into lesions. Fourth, viral latency is an unknown aspect of HPV that Gravitt points out, however, this model does not incorporate a latent period in the spreading process \cite{gravitt2011known}. Finally, we do not account for virion decay and instead focus on cumulative expulsion of virions from cells that have shed, or are considered dead. The assumptions of master equations are as follows: master equations define unique state changes that happen in continuous time. That being said, the number of such states is constrained by how computationally expensive it is to run the resulting system of equations. It is important to note that the master equations and method of moments are all exact, however, the extinction analysis is no longer exact since assumptions on the distributions are imposed.  \textit{The purpose of this model is to provide insight into epithelium dynamics modeling and the role of epithelium structure using the method of moments to reduce computational cost.} 

\subsection{\label{sec:me} Master equations}
Consider the general first order linear differential equation, defining the instantaneous rate of change for the discrete variable $y$,

\begin{align}\label{eq:example_ode}
    \frac{dy}{dt} + p(t)y = g(t).
\end{align}
The equations $p(t)$ and $g(t)$ define time-dependent processes \cite{boyce2020elementary}. When this framework is applied to population dynamics, where $y$ represents the population density at time $t$, we can solve for an exact solution. Due to its deterministic nature, solutions give a specific \textit{unique} population density at a time $t$. To transition from a deterministic framework to a stochastic framework, which accounts for the randomness of certain processes, while still using differential equation structure, we look to master equations. 

In this stochastic framework, the discrete variable is replaced by the probability associated with the state $y$, denoted by $C_{y}(t)$. For all states, the probability distribution incorporates stochasticity, which is beneficial for integrating uncertainty into this model. Master equations also substitute equations $p(t)$ and $q(t)$ for transition rates, $\omega_t$, which define a stochastic rate at which the state of the system changes. These stochastic rates can be thought of as the transfer of probability mass from one state to another \cite{haag2017modelling}. We define a general master equation as, 
\begin{equation} \label{eq:general_me}
    \frac{d}{dt}C_{y}(t) = - \sum_{z} \omega_{t}(y, z) C_{y}(t) +\sum_{z} \omega_{t}(z, y) C_{z}(t) ,
\end{equation}
which represents the change in the occupation probability or probability associated with state $y$. The left-hand summation  of Eq. (\ref{eq:general_me}) defines the subtraction of mass from the probability density, $C_{y}(t)$, meaning the probability mass is leaving state $y$ to any state $z$.  With the addition of probability mass shown in the right-hand summation, we notice probability mass moving from any state $z$ to $y$ \cite{haag2017modelling}. It is important to notice that the arbitrary state $y$ represents all information relevant to the state, and can therefore be a scalar or a multi-dimensional quantity. 

As an example, consider an arbitrary state which represents the scalar quantity $n$. This quantity tracks the number of infected cells in the system. The stochastic rates of transitions for this system are the infection rate, $\omega_{i}$ and the recovery rate, $\omega_{r}$. We define three master equations for $n = \{1,2,3\}$ as

\begin{align}
    \frac{d}{dt}C_{n=1}(t) &= - \omega_{i} C_{n=1}(t) - \omega_{r} C_{n=1}(t)  + \omega_{r} C_{n=2}(t),\\
    \frac{d}{dt}C_{n=2}(t) &= - \omega_{i} C_{n=2}(t) - \omega_{r} C_{n=2}(t)+ \omega_{i} C_{n=1}(t) + \omega_{r} C_{n=3}(t),\\
    \frac{d}{dt}C_{n=3}(t) &= - \omega_{i} C_{n=3}(t) - \omega_{r} C_{n=3}(t) + \omega_{i} C_{n=2}(t).
\end{align}

Now, we have a definition for the change in the probability density for $n = \{1,2,3\}$. From these equations we can determine the probability density at time $t$ for each $n$. The total number of states for this system is the range of $n$, meaning we have three possible states to track and have two degrees of freedom. Since a single dimension master equation model represents the probability distribution dynamics for $n$, we can extend this framework to a multi-dimensional master equations. The multi-dimensional master equations have the ability to represent population interactions. We define another quantity, $h$, representing the number of healthy cells. Therefore we can define $C_{n,h}(t)$ changing over time for $n=\{1,2\}$ and $h=\{1,2\}$ as 

\begin{align}
    \frac{d}{dt}C_{n=1, h=1}(t) &= - \omega_{i} C_{n=1, h=1}(t) - \omega_{r} C_{n=1, h=1}(t),\\
    \frac{d}{dt}C_{n=2, h=1}(t) &= - \omega_{i} C_{n=2, h=1}(t) - \omega_{r} C_{n=2, h=1}(t)+ \omega_{i} C_{n=1, h=2}(t),\\
    \frac{d}{dt}C_{n=1, h=2}(t) &= - \omega_{i} C_{n=1, h=2}(t) - \omega_{r} C_{n=1, h=2}(t) + \omega_{r} C_{n=2,h=1}(t),\\
    \frac{d}{dt}C_{n=2, h=2}(t) &= - \omega_{i} C_{n=2, h=2}(t) - \omega_{r} C_{n=2, h=2}(t).
\end{align}

By adding another cell, we alter where the probability mass is transitioning from, which showcases the population interaction possible with this method. The number of states also increases to four, providing three degrees of freedom, which is achieved by multiplying each quantity range by the other. We will use this multi-dimensional approach moving forward, and specify the types of transitional interactions between each quantity in the system. As shown in the examples, this paper defines a quantity as a unique cell type, therefore, the multi-dimensional system is renamed as a multi-cell-type system. Sections  \ref{sec:three} and \ref{sec:five} define a three-cell-type system, representing a three layer epithelium, and a five-cell-type system, representing a five layer epithelium.

\subsection{\label{sec:mom}Method of moments (MoM)}
From the small multi-cell-type example in Sec. ~\ref{sec:me}, we define a system with two cell types and see the number of equations to satisfy with a solution grows exponentially with the number of cells. For multi-cell-type models with large ranges, the state spaces grows quite large and this explicit method becomes computationally expensive. A simple solution to avoid solving for a large number of states is to derive the mean and variance from the probability distribution being defined. A mean value and variance description has been deemed adequate for describing large state spaces or populations \cite{haag2017modelling}. This process is not as computationally expensive as explicitly solving all the master equations, since there are only the chosen moments of each state and their interaction terms to track over time.

The mean for a given cell, $k$, in the cell range of 0 to $L$, $y = (y_{1}, y_{2}, y_{3}, \ldots, y_{L})$ is,
\begin{equation}\label{eq:general_mean}
    \bar{y}_{k}(t) = \sum_{y} y_{k}C_{y}(t) = \langle y_{k} \rangle _{t},
\end{equation} 
where $C_{y}(t)$ is the probability distribution for all states. The master equations solve for the change over these probability distributions, so using Eq. (\ref{eq:general_mean}), we define the equation for the change in the moment of a distribution as
\begin{equation}\label{eq:general_mom}
    \frac{d}{dt}\langle y_{k} \rangle_{t} = \sum_{y} y_{k}\frac{d C_{y}(t)}{dt}.
\end{equation}
Now, given a multi-cell-type system, we must consider higher order moments and interaction terms as 

\begin{equation}\label{eq:general_mom_higher_moments}
    \frac{d}{dt}\langle y^{\ell}_{k}y^{q}_{s} \rangle_{t} = \sum_{y} y^{\ell}_{k}y^{q}_{s}\frac{d C_{y}(t)}{dt}.
\end{equation}
From the first and second moment of state $y$, we define the variance as 

\begin{equation}\label{eq:variance}
    Var(y_k)_t = \langle y_{k}^{2} \rangle_{t} - \langle y_{k} \rangle^{2}_{t}. 
\end{equation}

\subsection{Extinction and non-extinction events \label{sec:extinction_math}}

The probability of extinction is essential for understanding the duration of an infection event, therefore, from the first two moments of any cell, we can derive the probability of extinction. We can do this from the assumption that the underlying distribution of a given average, $\langle y_k \rangle$, follows a zero-inflated geometric distribution. Kendall previously showed that the geometric distribution is a solution to the birth-death process \cite{kendall1948generalized} therefore, we use a geometric approximation of non-extinct trajectories to extract the probability of extinction from the method of moments. We redefine the first two moments as,

\begin{equation}\label{eq:geom_approx_first_moment}
\langle y \rangle  = \frac{1}{p} [1 - P(y = 0)],
\end{equation}

\begin{equation}\label{eq:geom_approx_sec_moment}
    \langle y^2 \rangle = \frac{2-p}{p^{2}}[1 - P(y = 0)],
\end{equation}
where $P(y = 0)$ is the probability of extinction for cell $y$ and $\frac{1}{p}$ and $\frac{2-p}{p^{2}}$ are the first and second moment of the geometric distribution for non-extinct states. Solving for $p$ from Eq. (\ref{eq:geom_approx_first_moment}), we define

\begin{equation}\label{eq:geom_approx_p}
    p = \frac{[1 - P(y = 0)]}{\langle y \rangle}.
\end{equation}
Now, substituting Eq. (\ref{eq:geom_approx_p}) into Eq. (\ref{eq:geom_approx_sec_moment}), we can solve for the probability of extinction as  

\begin{equation}\label{eq:prob_extinction}
     P(y = 0) = 1 -\frac{2 \langle y \rangle^{2}}{\langle y^{2}\rangle + \langle y \rangle},
\end{equation} 
which provides a solution for the probability of extinction using the first two moments of the distribution of associated with cell $y$. In Sec. ~\ref{sec:results}, details on transient infections defined by the probability of extinction will be discussed. 
Now, when solving for the probability of extinction, we inherently solve for the probability of non-extinction as well, 
\begin{equation}\label{eq:prob_non-extinction}
     1 - P(y = 0) = \frac{2 \langle y \rangle^{2}}{\langle y^{2}\rangle + \langle y \rangle}.
\end{equation} 
From the probability of non-extinction, we can estimate the geometric distribution parameter with a specific cell from Eq. ~(\ref{eq:geom_approx_p}),
\begin{equation}\label{eq:solving_p_with_both_moments}
    p = \frac{2\langle y \rangle }{\langle y^{2} \rangle + \langle y \rangle },
\end{equation}
thus estimating the mean and variance of a cell count for a persistent infection. This provides valuable insights for the average basal cells from a persistent infection, which will be discussed in Sec. ~\ref{sec:results}.

\subsection{Viral load \label{sec:viral_load}}
The viral load metric in an individual is the product of infected cells shedding from the epithelium. When a cell is shed, it disperses virions or copies of HPV that can go on to infect others. As pointed out in Sec. \ref{sec:intro}, there are many uncertainties around the progression of an HPV infection, one of which is viral load \cite{gravitt2011known, ryser2017mechanistic}. It is known that when a cell becomes infected by HPV, there is a genome replication that occurs in the basal cell layer, resulting in 50-100 viral copies in the cell \cite{doorbar2007papillomavirus, stanley2012epithelial}. As infected cells differentiate and move up the cell layers, more viral copies are produced within the cell. The number of copies varies according to the genotype, however, we set the number of virions to 1,000 as a proof of concept. We determine viral load output from the MoM by focusing on the moments associated with the final cell type, dead cells. We determine the first and second moment of the dead cells, then apply the virion shed value for each dead cell, producing the expected total number of virions released over time.   

In the future we can aim to give a general distribution for the viral copies per cell \cite{garner2002hpv, swan1999human}. When a cell is shed from the epithelium, it is estimated between $50 - 10^{4}$ viral copies are expelled.  \cite{stanley2012epithelial, moody2017mechanisms, garner2002hpv, swan1999human}. Due to this wide range for viral copies shed out of the system, we can apply a specified distribution for the previously mentioned range. This distribution could be altered according to new research or other assumptions, for example, to include more stochasticity, the distribution over the given range could be uniform. 

\subsection{Stochastic simulation algorithm \label{sec:sims}}
Without data for these cellular dynamics, we simulate the stochastic process of an infection in the cervical epithelium with the Gillespie stochastic simulation algorithm \cite{gillespie1977exact}. This algorithm establishes a continuous-time simulation, which tracks the events that occur as time passes. These event-driven simulations are conducted by distinguishing the rates of each event, then assigning each rate to its respective exponential distribution. Drawing from all the exponential distributions provides the next time to all specific events. Whichever time is the closest to the current time means the associated event will occur. New events based off the event that just occurred are placed in the queue, since one event triggers a future event. This continues until time runs out or all cell types are equal to 0, excluding the dead cells. For the purposes of this project, we set a maximum time of 750 days to occur for 50,000 simulations.

\section{Application to structured epithelium dynamics \label{sec:applications}}

\subsection{\label{sec:three}Three-cell-type system}

\begin{figure}
    \centering
    \includegraphics[width = 0.85\linewidth]{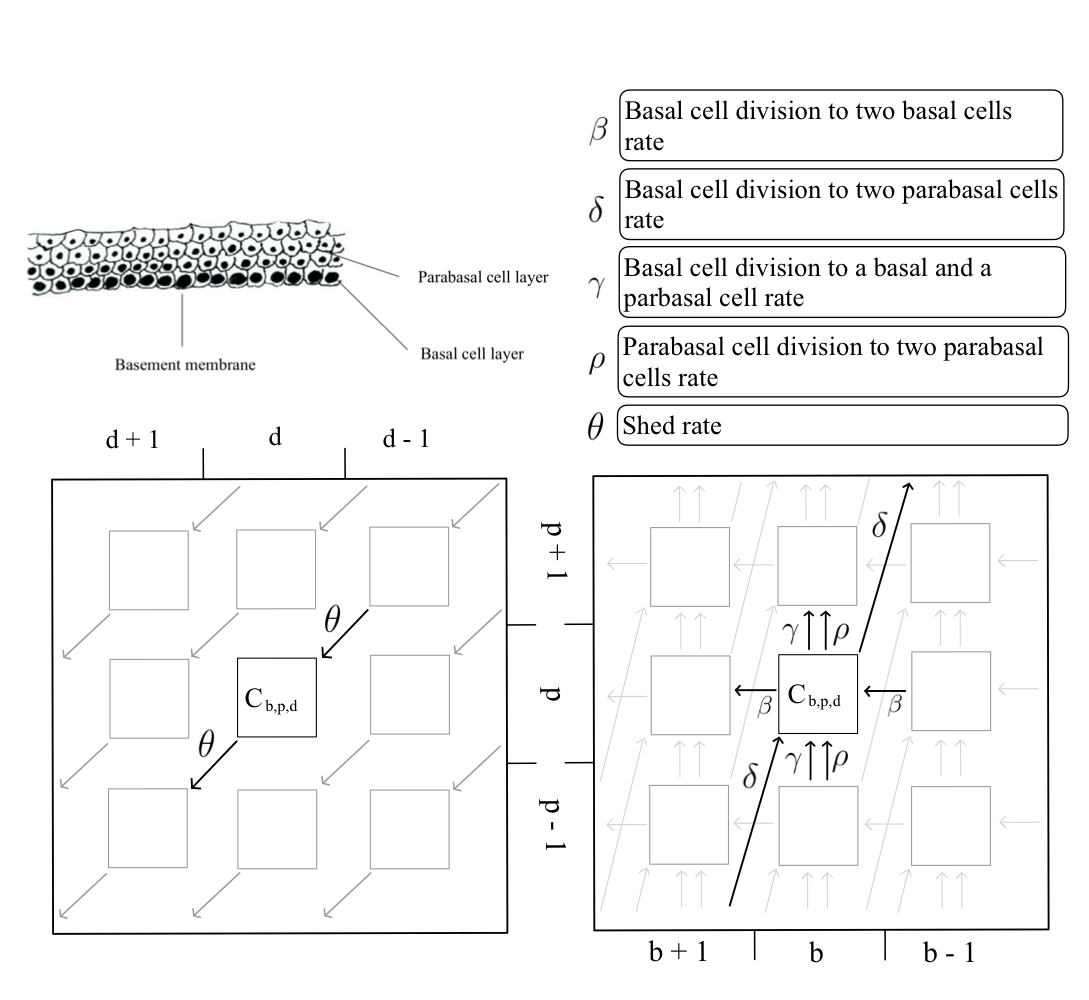}
    \caption{\textbf{Three-cell-type model schematic:} Each state of this system is fully specified by the number of infected basal cell $b$, the number of infected parabasal cells $p$, and the number of infected dead cells shed $d$. The general dynamics are shown with arrows illustrating the probability mass transitions between states. A state change is made when the infected basal cell count changes, the infected parabasal cell count changes, or the dead cell count changes in the system\footnote{Cell image reprinted from Colposcopy and Treatment of Cervical Precancer,  IARC technical publication No. 45, Walter Prendiville and Rengaswamy Sankaranarayanan, Anatomy of the uterine cervix and the transformation zone, Page 14, 2017 (2017) \cite{prendiville2017colposcopy}.}.}
    \label{fig:three_state_schematic}
\end{figure}

For the first iteration of this model, we will only focus on the first two layers of the cervical epithelium, the basal and parabasal cell layers. The biological reason for focusing on the bottom two layers first is that after menopause, the cells in the cervix do not mature past the parabasal cell layer. This results in a thinner cervical epithelium \cite{prendiville2017colposcopy}. After an initial basal cell is infected, set as an initial condition, there are three types of basal cell divisions that can accumulate more infected cells in the epithelium. An infected basal cell can divide into two more infected basal cells, divide into an infected basal cell and an infected parabasal cell, or divide into two infected parabasal cells. These divisions happen with rates of $\beta$, $\gamma$, and $\delta$ respectively. Once an infected parabasal cell enters the system, it has the potential to divide into two more infected parabasal cells, or differentiate, which allows the cell to shed from the epithelium. Each of these processes occur with respective rates of $\rho$ and $\theta$. The resulting system is therefore fully specified by the number of infected basal cells, $b$, the number of infected parabasal cells, $p$, and the number of infected dead cells shed, $d$. The epithelial dynamics between these three types of cells are depicted in Fig. \ref{fig:three_state_schematic}. The parameter values are shown in Table \ref{table:threestate_parameters}, which are used in the MoM equations and simulations.

The master equation derived from Fig.~\ref{fig:three_state_schematic} tracks three quantities: $b$, $p$, and $d$. The general probability distribution over all possible states $\{b, p, d\}$ is defined as $C_{b,p,d}(t)$, which evolves in time according to,

   \begin{align} \label{eq:two_state_explicit_me}
     \frac{d}{dt}C_{b,p,d}(t) = & (b-1)\beta C_{b-1,p,d} + (p+1)\theta C_{b,p+1,d-1} + ((p-1)\rho  + b\gamma) C_{b,p-1,d} + (b+1)\delta C_{b+1,p-2,d} \\ \nonumber
     & - [b\beta + b\gamma + b\delta + p\rho + p\theta]C_{b,p,d}. 
    \end{align}

As explained previously, if we track our counting variables $b$, $p$ and $d$ up to some large integer $N$, we are then dealing with a system of $N^3$ equations which can become unwieldy when dealing with realistic infection sizes. Thus we move to the method of moments as a computationally inexpensive alternative. This work focuses on the first and second moments of all the cell variables. The derivations of these moments follow from Eq. (\ref{eq:general_mom}). The exact equations are given by

\begin{equation*}
\begin{aligned}[c]
\frac{d}{dt} \langle b\rangle & = (\beta - \delta) \langle b \rangle \nonumber \\
\frac{d}{dt} \langle b^{2}\rangle & = 2(\beta - \delta)\langle b^{2} \rangle  + (\beta + \delta)\langle b \rangle \nonumber\\
   \frac{d}{dt} \langle p\rangle & = (\rho - \theta)\langle p \rangle + (2\delta + \gamma)\langle b \rangle \nonumber\\
   \frac{d}{dt} \langle p^{2}\rangle & = (\theta + \rho)\langle p \rangle  + (2\rho - 2\theta)\langle p^{2} \rangle + (\gamma + 4\delta)\langle b \rangle + (2\gamma + 4\delta)\langle bp \rangle \nonumber \\
   \frac{d}{dt} \langle bp\rangle & = (\beta - \theta + \rho - \delta)\langle bp \rangle  + (\gamma + 2\delta)\langle b^{2} \rangle -2\delta \langle b \rangle \nonumber \\
\end{aligned}
\qquad 
\begin{aligned}[c]
   \frac{d}{dt} \langle d\rangle & = \theta \langle p \rangle \nonumber\\
   \frac{d}{dt} \langle d^{2}\rangle & = \theta \langle p \rangle + 2\theta \langle pd \rangle \nonumber \\
   \frac{d}{dt} \langle pd\rangle & = (\rho - \theta) \langle pd \rangle + \theta \langle p^{2} \rangle - \theta \langle p \rangle + (\gamma + 2\delta)\langle bd \rangle \nonumber \\
   \frac{d}{dt} \langle bd\rangle & = \beta \langle b \rangle + \theta \langle bp \rangle - \delta \langle bd \rangle.
\end{aligned}
\end{equation*}

\begingroup

\renewcommand{\arraystretch}{1.5}
\begin{table}[h!]

\begin{tabular}{|c | l | c | c|} 
 \hline
 & Process definition & Rate [1/days]  & Reference\\ [0.5ex] 
 \hline
 $\beta$ & Basal cell to two basal cells division & 0.0034 & \cite{beneteau2021episome, murall2019epithelial, clayton2007single} \\ 
  \hline
 $\delta$ & Basal cell to two parabasal cells division  & 0.0024 &  \cite{beneteau2021episome, murall2019epithelial,clayton2007single} \\
  \hline
 $\gamma$ & Basal cell to one basal cell and one parabasal cell division & 0.0252 &  \cite{beneteau2021episome, murall2019epithelial, clayton2007single}\\
  \hline
 $\rho$ & Parabasal cell to two parabasal cells division & 0.0312 &  \cite{murall2019epithelial, clayton2007single} \\
  \hline
 $\theta$ & Parabasal cell shed & 0.67 & \cite{murall2019epithelial}\\ [1ex] 
 \hline
\end{tabular}
\caption{Parameter values for the processes occurring on the system are used in both the analytical model and the simulations for the three-cell-type system. All rates are measured in units of 1/days.}
\label{table:threestate_parameters}
\end{table}
\endgroup

Numerical solutions to these equations give enough information to derive the distribution's mean and standard deviation for each cell at time $t$. These exact analytical results can the be compared to the outcome of the simulation process defined in Sec. ~\ref{sec:sims} starting from a single infected basal cell. The simulation then tracks the history of each cell type count over the 750 days.

\subsection{\label{sec:five} Five-cell-type system}
\begin{figure}[h]
    \centering
    \includegraphics[width = 0.95\linewidth]{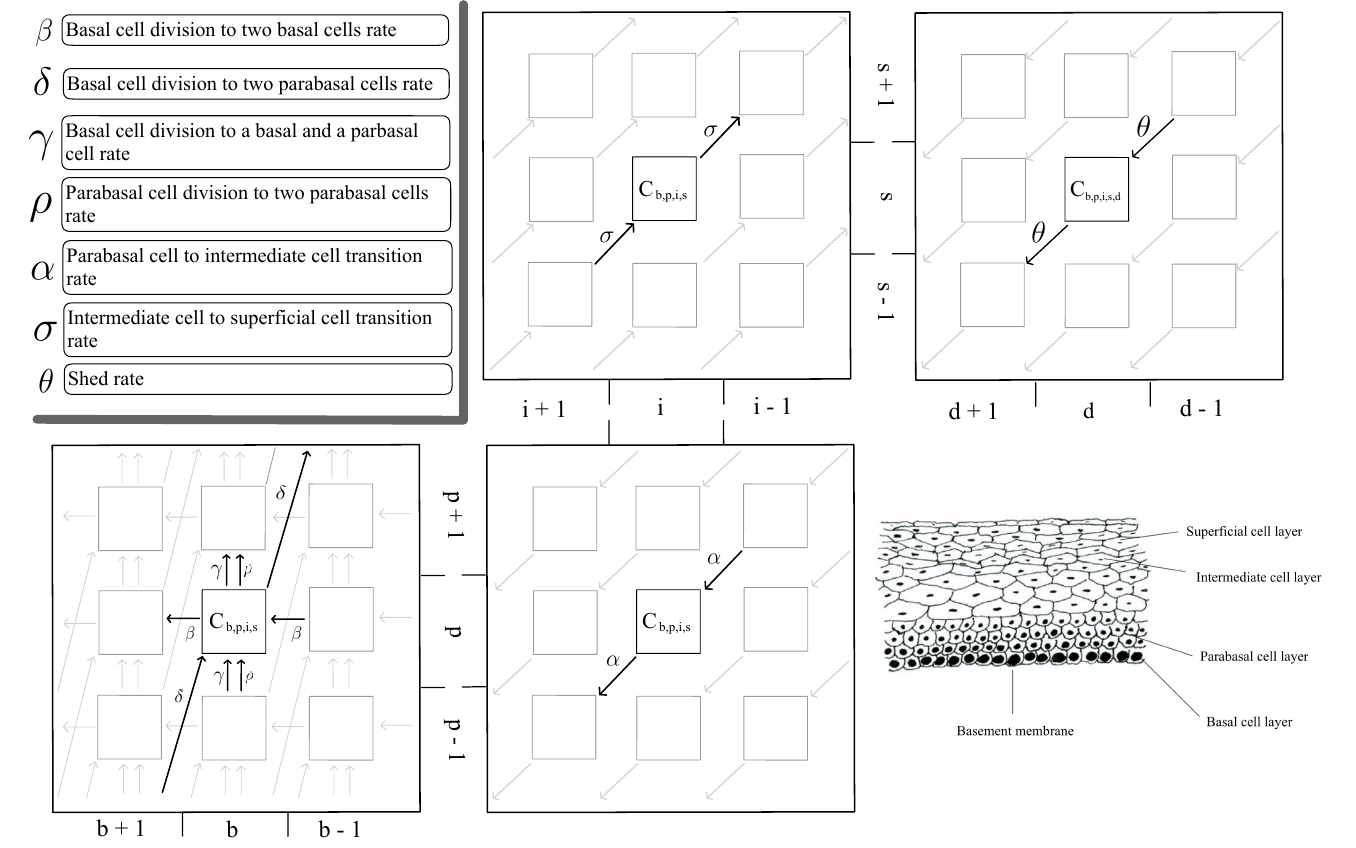}
    \caption{\textbf{Five-cell-type model schematic:} Similar to the three-cell-type system, the general dynamics for the five-cell-type system are shown with arrows illustrating the probability mass transitions between states. A state change is made when any of the infected cell counts or the dead cell count change in the system. With the additional cell types, we see the differing dynamics from parabasal cells to intermediate and superficial cells \footnote{Cell image reprinted from Colposcopy and Treatment of Cervical Precancer,  IARC technical publication No. 45, Walter Prendiville and Rengaswamy Sankaranarayanan, Anatomy of the uterine cervix and the transformation zone, Page 14, 2017 (2017) \cite{prendiville2017colposcopy}}.}
    \label{fig:five_state_schematic}
\end{figure} 

The second iteration of this model adds the dynamics of the two upper layers in the cervical epithelium, the intermediate, $i$, and superficial, $s$, cell layers. The general probability distribution of infected cells for each cell type is changed to $C_{b,p,i,s,d}(t)$ to incorporate the extra layers. Similar to Sec. ~\ref{sec:three}, once an initial basal cell is infected, infected cells are introduced to other layers through divisions or transitions. Previously, we also described how infected basal cells can divide into two infected basal cells, an infected basal cell and parabasal cell, or two infected parabasal cells. Remember, these divisions occur at rates of $\beta$, $\gamma$, and $\delta$ respectively. Now, the intermediate and superficial cells do not result from division dynamics, rather, they populate through transition dynamics. Therefore, the dynamics of the infected parabasal cells diverge from the three-cell-type model dynamics. Instead, an infected parabasal cell either divides into two infected parabasal cells with rate $\rho$, or transitions into an infected intermediate cell with rate $\alpha$. Furthermore, intermediate cells cannot divide, so, the only event that can occur is for the cell to transition up to an infected superficial cell with rate $\sigma$. Lastly, an infected superficial cell will die and shed any virions with rate $\theta$ \cite{prendiville2017colposcopy}. The interactions between subsequent layers are shown in Fig. \ref{fig:five_state_schematic}. The parameter values for each process rate are defined in Table \ref{table:fivestate_parameters}.

The master equation derived from Fig.~\ref{fig:five_state_schematic} details the flow in and out of each state. The equation is as follows, 
   \begin{align}\label{eq:five_state_explicit_me}
     \frac{d}{dt}C_{b,p,i,s,d}(t) = & (b-1)\beta C_{b-1,p,i,s,d} + ((p-1)\rho + b\gamma)C_{b,p-1,i,s,d} + (b+1)\delta C_{b+1,p-2,i,s,d} \\ \nonumber
     &  + (p+1)\alpha C_{b,p+1,i-1,s,d} + (i+1)\sigma C_{b,p,i+1,s-1,d} + (s+1)\theta C_{b,p,i,s+1,d-1}  \\ \nonumber
     & - (b\beta + p\rho + b\gamma + b\delta + p\alpha + i\sigma + s\theta ) C_{b,p,i,s,d} \; .
    \end{align} 

Similar to the three-cell-type system, we derive the first and second moments for each of the five cell types, along with the pairwise interaction terms. The five-cell-type simulation follows the same process as described in Sec. \ref{sec:three}, except the $i$ and $s$ states are also tracked. All exact MoM equations for the five-cell-type system are given by  

\begin{equation*}
\begin{aligned}[c]
\frac{d}{dt} \langle b\rangle & = (\beta - \delta) \langle b \rangle \nonumber\\
   \frac{d}{dt} \langle p\rangle & = (\rho - \alpha)\langle p \rangle + (\gamma + 2\delta)\langle b \rangle \nonumber\\
   \frac{d}{dt} \langle i\rangle &= \alpha \langle p \rangle -\sigma \langle i \rangle \nonumber\\
   \frac{d}{dt} \langle s \rangle &= \sigma \langle i \rangle -\theta \langle s \rangle \nonumber\\
   \frac{d}{dt} \langle d \rangle &= \theta \langle s \rangle \nonumber\\
   \frac{d}{dt} \langle b^{2}\rangle & = 2(\beta - \delta)\langle b^{2} \rangle  + (\beta + \delta)\langle b \rangle \nonumber \\
   \frac{d}{dt} \langle p^{2}\rangle & = (\rho + \alpha)\langle p \rangle  + (2\rho - 2\alpha)\langle p^{2} \rangle + (2\gamma + 4\delta)\langle bp \rangle + (\gamma + 4\delta)\langle b \rangle \nonumber \\
   \frac{d}{dt} \langle i^{2} \rangle & = \alpha \langle p \rangle + 2\alpha \langle pi \rangle -2\sigma \langle i^{2} \rangle + \sigma \langle i \rangle \nonumber \\
   \frac{d}{dt} \langle s^{2} \rangle & = \sigma \langle i \rangle + 2\sigma \langle is \rangle -2 \theta \langle s^{2} \rangle + \theta \langle s \rangle \nonumber \nonumber \\
   \frac{d}{dt} \langle d^{2} \rangle & = 2 \theta \langle sd \rangle + \theta \langle s \rangle \nonumber 
\end{aligned}
\qquad 
\begin{aligned}[c]
   \frac{d}{dt} \langle bp\rangle & = (\beta + \rho - \delta - \alpha)\langle bp \rangle  + (\gamma + 2\delta)\langle b^{2} \rangle -2\delta \langle b \rangle \nonumber\\
   \frac{d}{dt} \langle bi \rangle & = (\beta - \delta - \sigma) \langle bi \rangle + \alpha \langle bp \rangle \nonumber\\
   \frac{d}{dt} \langle bs \rangle & = (\beta - \delta - \theta) \langle bs \rangle + \sigma \langle bi \rangle \nonumber \\
   \frac{d}{dt} \langle bd \rangle &= (\beta - \delta)\langle bd \rangle + \theta \langle bs \rangle \nonumber \\
    \frac{d}{dt} \langle pi \rangle & = (\rho - \alpha - \sigma)\langle pi \rangle + (\gamma + 2\delta)\langle bi \rangle + \alpha \langle p^{2} \rangle - \alpha \langle p \rangle \nonumber \\
    \frac{d}{dt} \langle ps \rangle & = (\rho - \theta - \alpha) \langle ps \rangle + (\gamma + 2\delta) \langle bs \rangle  + \sigma \langle pi \rangle \nonumber \\
    \frac{d}{dt} \langle pd \rangle &= (\rho - \alpha)\langle pd \rangle + (\gamma + 2\delta) \langle bd \rangle + \theta \langle ps \rangle \nonumber\\
    \frac{d}{dt} \langle is \rangle & = (-\theta - \sigma) \langle is \rangle + \alpha \langle ps \rangle + \sigma \langle i^{2} \rangle - \sigma \langle i \rangle \nonumber\\ 
    \frac{d}{dt} \langle id \rangle & = \theta \langle is \rangle - \sigma \langle id \rangle + \alpha \langle pd \rangle \nonumber\\ 
    \frac{d}{dt} \langle id \rangle & = \theta \langle is \rangle - \sigma \langle id \rangle + \alpha \langle pd \rangle \nonumber\\ 
    \frac{d}{dt} \langle sd \rangle & = \theta \langle s^{2} \rangle - \theta \langle sd \rangle - \theta \langle s \rangle + \sigma \langle id \rangle \nonumber
\end{aligned}
\end{equation*}

The exponential distributions for the time to next event considers the additional processes of parabasal cells moving to the intermediate cell layer, intermediate cells moving to the superficial cell layer. Finally, superficial cells shed out of the epithelium and add to the cumulative number of dead cells.

\begingroup

\renewcommand{\arraystretch}{1.5}
\begin{table}[ht]
\begin{tabular}{|c | l | c | c|} 
 \hline
 & Process definition & Rate [1/days]  & Reference\\ [0.5ex] 
 \hline
 $\beta$ & Basal cell to two basal cells division & 0.0034 & \cite{beneteau2021episome, murall2019epithelial, clayton2007single} \\ 
  \hline
 $\delta$ & Basal cell to two parabasal cells division  & 0.0024 &  \cite{beneteau2021episome, murall2019epithelial,clayton2007single} \\
  \hline
 $\gamma$ & Basal cell to one basal cell and one parabasal cell division & 0.0252 &  \cite{beneteau2021episome, murall2019epithelial, clayton2007single}\\
  \hline
 $\rho$ & Parabasal cell to two parabasal cells division & 0.0312 &  \cite{murall2019epithelial, clayton2007single} \\
  \hline
  $\alpha$ & Parabasal cell differentiating and moving to the intermediate cell layer  & 0.4 & \cite{murall2019epithelial} \\ 
  \hline
  $\sigma$ & Cell moving from the intermediate cell layer to the superficial cell layer  & 0.4 & \cite{murall2019epithelial} \\ 
  \hline
 $\theta$ & Superficial cell shed & 0.67 & \cite{murall2019epithelial}\\ [1ex] 
 \hline
\end{tabular}
\caption{Parameter values used in both the analytical model and the simulations for the five-cell-type system. All rates are measured in units of 1/days.}
\label{table:fivestate_parameters}
\end{table}
\endgroup

\section{Results \label{sec:results}}

\begin{figure*}[ht]
    \centering
    \includegraphics[width = 0.85\linewidth]{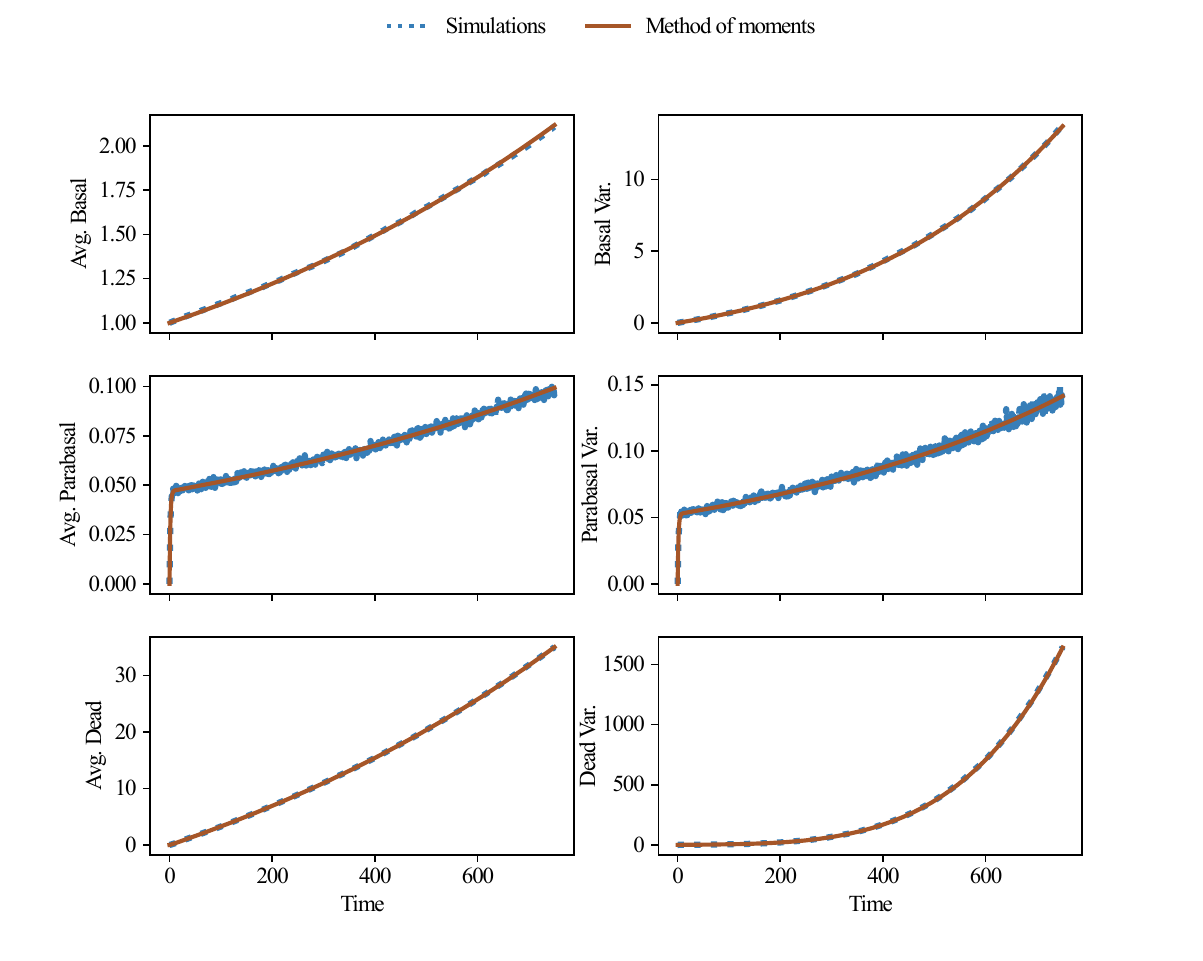}
    \caption{\textbf{Average and variance evolution of the three-cell-type system:} The averages and variances from the MoM equations for the basal, $b$, parabasal, $p$, and dead, $d$, cells over 750 days (solid lines) are validated by the simulations (dotted lines). 50,000 simulations were used to compute the mean and variance for each cell type. The initial conditions for the MoM and simulations set the first and second moments of $b$ equal to 1, while the other moments are set to 0.}
    \label{fig:three_mom_sims}
\end{figure*}

Section \ref{sec:three}  defines the analytical moments and the event-driven simulations for the three-cell-type system over time. Figure \ref{fig:three_mom_sims} shows the results and exactitude of the MoM equations compared to the simulations. As time goes on, the average number of infected basal cells grows slowly to 2 by 750 days. This indicates that the basal cell division rate is not large enough to double the average until after 750 days. The variance of the basal cells gradually increases over time as well, illustrating that the longer the infection persists, the more variable the basal cell counts can be. Turning to the average count of infected parabasal cells, the average jumps from 0, implying immediate divisions of infected basal cells to infected parabasal cells. The change in the infected parabasal cell average is not as large as the change in the infected basal cell average, but intuitively we attribute this to the rate of shedding being larger than all other rates in the system. It is noteworthy that the simulated average of infected parabasal cells is more variable over time, which is a result of the small values for the average infected parabasal cells. This scale could also have an effect on the variance of the parabasal cells, which is small relative to the other cell-type variances. Moving on to the dead cells, there is a steady increase in the average over the time period, similar to the average basal cells. However, the variance of the dead cells grows at a fast rate, likely due to the large shed rate and the variability from the average parabasal cell counts. 

\begin{figure}[ht]
    \centering
    \includegraphics[width = \linewidth]{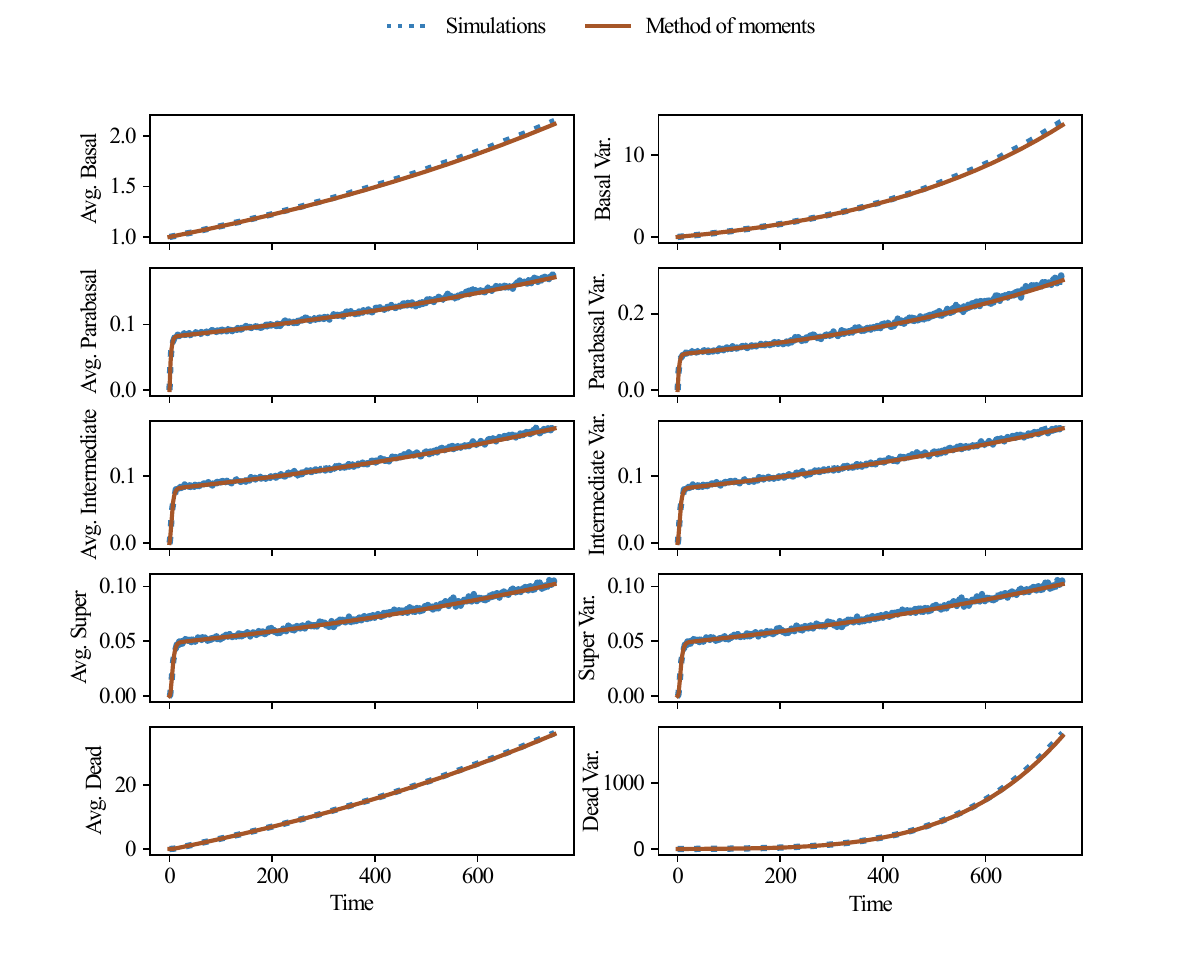}
    \caption{\textbf{Average and variance evolution of the five-cell-type system:} From the first and second moments of the basal, $b$, parabasal, $p$, intermediate, $i$, superficial, $s$, and dead, $d$, cells over 750 days, we derive the mean and variance (solid lines) for the five-cell-type system. The initial conditions are the same as the three-cell-type system, where the first and second moments for the basal cells are set to 1. All other moments are set to 0 at the start. We validate these analytical results with simulations (dotted lines). Similar to the three-cell-type system, 50,000 simulations were used for this validation.}
    \label{fig:five_mom_sims}
\end{figure}

With the intermediate and superficial cell layers to consider, the same analyses conducted for the three-cell-type system are conducted for the five-cell-type system. Figure \ref{fig:five_mom_sims} shows that with two more cell types, the simulations still validate the MoM results. The figure provides similar insights as Fig. \ref{fig:three_mom_sims} for the basal cell average and variance. On the other hand, the average and variance for the parabasal cells exhibit larger quantities over time. Turning to the average count of intermediate and superficial cells, both exhibit a similar trend as the average parabasal cell count. One major difference is the average count for parabasal and intermediate cells are larger than the average count for superficial cells. While the magnitude of the variances differ for the parabasal, intermediate, and superficial cells, each have a slow growing variance. This slow growth and variability in simulations is a result of the small averages of each cell type. Sec ~\ref{sec:viral_load} compares the dead cell averages in terms of viral load to highlight the differences between the two systems.

By accurately modeling the mean and variances of each cell type, we have the ability to understand the distribution of cell type counts over time. Furthermore, these distributions can encapsulate the likelihood of various disease progressions, for example viral load output, shown in Section \ref{sec:viral_load}.  With the MoM being exact for all cell types, we assert that this method has the potential to derive probabilities of extinction, persistent infection indicators and average cumulative virion counts to inform infectiousness of an individual in population models.

\subsection{Extinction probability \label{sec:extinct_prob}}
\begin{figure}[ht]
    \centering
    \includegraphics[width = \textwidth]{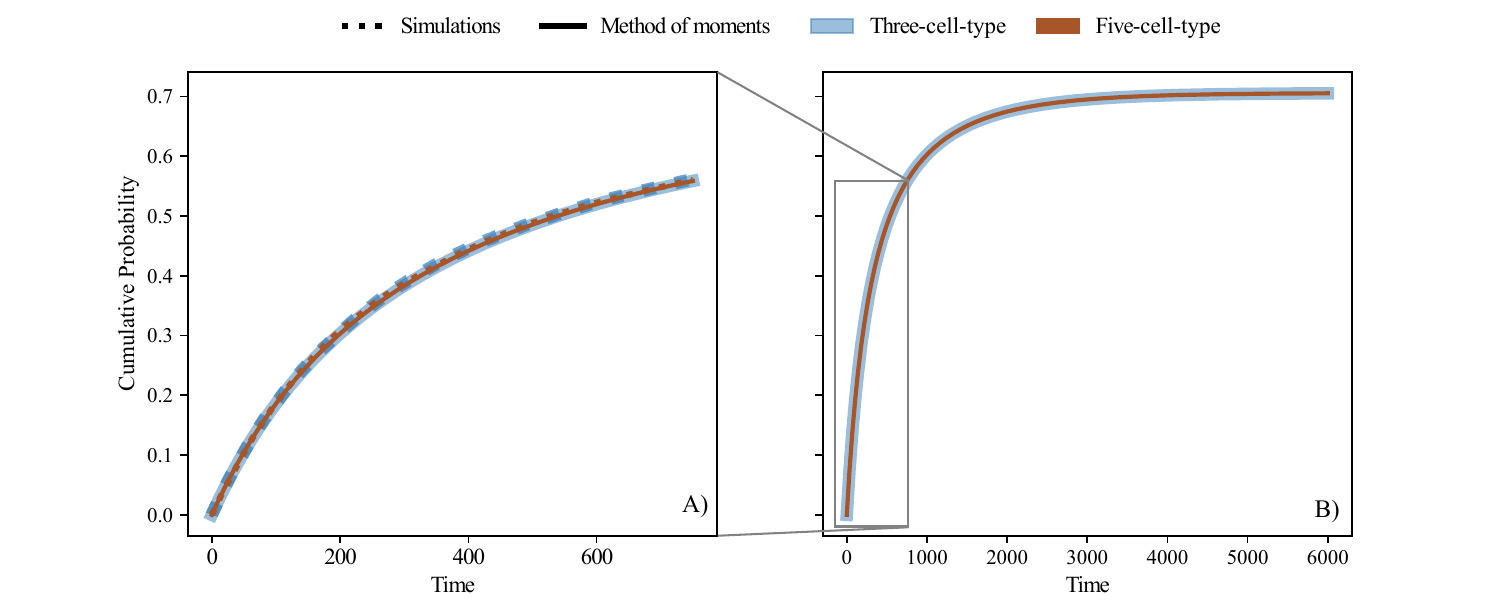}
    \caption{\textbf{Cumulative probability of extinction of the basal cells for the three and five-cell-type system:} In panel A, 50,000 simulations (dotted lines), over a 750 day time period, validate the probability of extinction derived from the first and second moments (solid lines) of the basal cells for the three (blue) and five-cell-type (red) system respectively. To derive the probability of extinction, we assume a zero-inflated geometric distribution, where we define the probability of extinction in terms of the first two moments. Panel B exhibits the probability of extinction after 6,000 days. Evaluating the probability of extinction past 750 days showcases the probability plateauing to 0.7 after 3,000 days. }
    \label{fig:prob_extinct}
\end{figure}

From the biological mechanisms of the epithelium, once the infected basal cell value is equal to 0, the parabasal cells will divide and eventually shed, eradicating the infection. Therefore, when the infected basal cell count reaches the zero-th column of Fig. \ref{fig:three_state_schematic} and \ref{fig:five_state_schematic}, this marks an extinction event. As shown in Sec. \ref{sec:extinction_math}, the probability of the infected basal cell average equaling 0, from our assumed zero-inflated geometric distribution, is equal to the probability of extinction. Remember this is not an exact result since we cannot assume a geometric approximation to be exact for all times. The analytical derivation is validated with simulations in panel A of Fig. \ref{fig:prob_extinct} for both modeled systems. We note that the differing systems do not affect the results of the extinction probabilities due to consistent basal cell dynamics, as shown in panels A and B. From the efficient MoM results, we are able to provide extinction probabilities for varying time periods. For example, panel A illustrates the probability of the average basal cells going extinct is over 50\% at 750 days. Extending past the 750 day mark in panel B, the probability of extinction levels out around 70\%. This means, after 6,000 days, there is a 70\% chance the infection has cleared out of the basal layer. While the likelihood of clearing an infection is high, there is still a chance of persistence, which is discussed in Sec. ~\ref{sec:persistent}.

\subsection{Persistent infections\label{sec:persistent}}

\begin{figure}[ht]
    \centering
    \includegraphics[width = 0.85\linewidth]{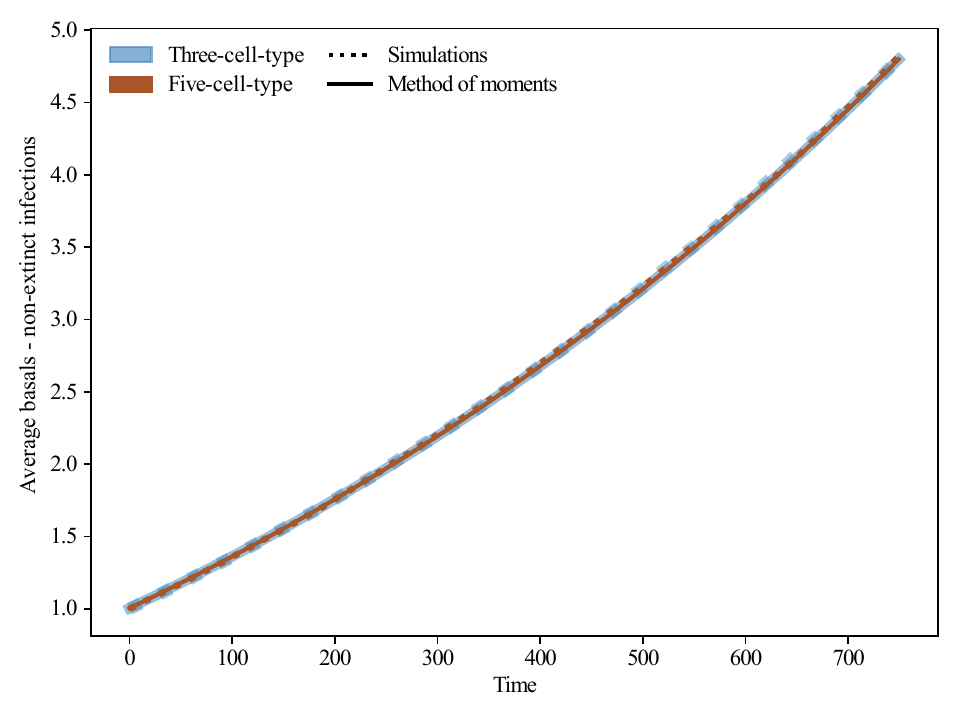}
    \caption{\textbf{Average basal cells of persistent infections for the three and five-cell-type system:} The non-extinct average basal cell simulation results (dotted lines) validate the MoM results (solid lines). The averages grow faster than those that include extinction events for both the three (blue) and five-cell-type (red) systems.}
    \label{fig:persistent}
\end{figure}

The dynamics of persistent infections are necessary for understanding the long term effects of an HPV infection. Therefore, we show the average basal cells without extinctions in Fig. ~\ref{fig:persistent} for both the three and five-cell-type system. Simulations excluding extinction events validate the MoM results for the average number of basal cells. Comparing the average basal cells in Figs. ~\ref{fig:three_mom_sims} and \ref{fig:five_mom_sims}, the average basal cells for non-extinct infections are much larger. Consequently, the larger this average is over time, the longer the infection could persist past this point. Moreover, these findings estimate the severity of 30\% of HPV infections that will persist over time. Persistent infections not only affects the infectiousness of an individual, but also their risk for cancer. The progression to cancer and infectiousness play vital roles in the disease spread and mortality rate in population-level models.

\subsection{Cumulative virions \label{sec:cumu_virions}}
\begin{figure}[ht]
    \centering
    \includegraphics[width = 0.85\linewidth]{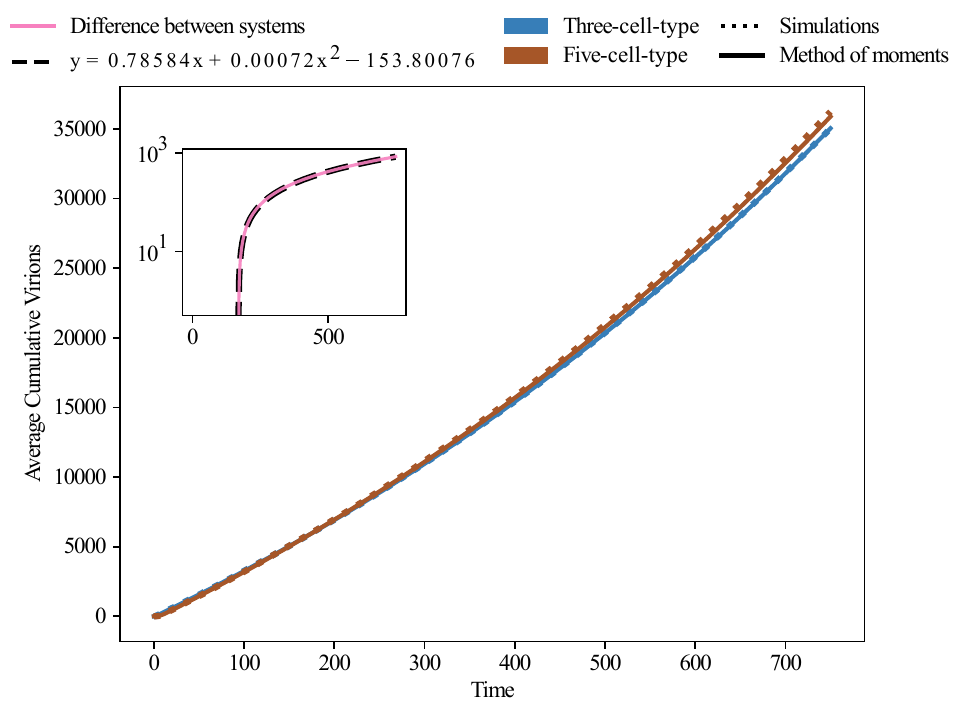}
    \caption{\textbf{Average cumulative virions shed for the three and five-cell-type system:} The average cumulative virion count from simulations (dotted lines) validate the MoM results (solid lines). The average cumulative virions shed is defined by 1,000 virions shed per dead cell. The difference in average cumulative virion count for the three (blue) and five-cell-type (red) system is highlighted in the inset plot (pink solid line), with a quadratic fit (black dashed line). }
    \label{fig:viral_load}
\end{figure}

Whether an infection is transient or persistent, we establish the average cumulative virions resulting from the epithelial dynamics. By assuming a certain number of virions shed from each dead cell, we achieve Fig. \ref{fig:viral_load}. This figure validates the MoM with simulations after a constant multiplier is applied to the average dead cell count for the three and five-cell-type system. The average cumulative virion count steadily grows as time goes on, ending at a little over 35,000 virions expelled at 750 day mark for the five-cell-type system. The two systems begin to significantly diverge between 150-250 days. Furthermore, the inset plot shows the difference between the cell-type systems over time (pink solid line). For comparison, a quadratic fit is also plotted in the inset (black dashed line), showing the difference follows a quadratic trend for the 750 day period. As time passes, the larger the gap is between the two systems, exhibiting the important role the extra layers have in harboring more infected cells. Since the importance of the extra layers depend on a person's age \cite{prendiville2017colposcopy}, our results suggest that different approximations for an individual's level of infectiousness over time should be used based on their age to better inform population models.

\section{\label{sec:discuss}Discussion}
While population level HPV models help inform persistent disease spread, within-host models have the ability to add individual-level stochasticity and heterogeneity to those population models. This is especially true since the mechanisms of cell division by which HPV infections propagate through an epithelium are well known. However, other less known factors such as clearance and role of epithelium layers must also be considered. With the uncertainty in the outcomes of an HPV infection preventing accurate modeling, it is imperative to leverage the knowledge of cell division dynamics to infer important infection dynamics. 

In this paper, we have demonstrated how master equations can track the probability distributions associated with count of infected cells in a cell layer. For each system, we define a generalized governing master equation for the probability of being in an arbitrary state. While the solutions from these master equations are exact, the larger the system, the more computationally expensive it is to solve. To avoid this, we can instead simply track the statistical moments of the full state distribution using the method of moments. Rather than tracking a full distribution for every cell type over time, these moments provide an accurate and efficient way to study the probabilistic nature of HPV infection progression. Moreover, we can still use the derived moments to compute the probability of extinction, average basal cell count for persistent infections, and average cumulative virions over time. In doing so, we tested the effects of adding structure to the epithelium in the form of extra cell layers whose importance are known to depend on age. We found that while this extra structure does not affect the probability of establishment (or extinction) of a new infection, it does affect long-term shedding rates of virions in persistent infections.

Some considerations for future work are incorporating reinfections, and within-host spatial tracking. For simultaneous infections, it would be important to allow current infections to reinfect the host stochastically, or introduce another transmission event to add to the viral load burden. This consideration could result in a change to the moments of each cell-type distribution, meaning the probability of extinction could go down drastically depending on the time of the reinfection event while the non-extinct average basal cell count could increase. This compounding effect could spike the average cumulative virions and change the level of infectiousness for the individual. Finally, incorporating a spatial component to this model could account for simultaneous infections merging to form a larger infected area. Consequently, this would account for possible lesion grades and treatments to rid an individual of the infection. 

While this framework informs HPV infection growth, epithelium infection progressions do not only show up in a cervical epithelium. This model and its outcomes can be applied to other parts of the body as well. Leveraging this framework for not only other parts of the body but for other diseases could benefit modelers that struggle with individual-level disease knowledge. The generalisability of master equations and the method of moments for infection progressions through cell division dynamics provide a powerful framework that can aid population-level models.

The reasons to model span from understanding population-level disease dynamics to mitigating outbreaks, but there are large assumptions placed on modeling diseases that have a number of unknown aspects. Using the information on the within-host mechanisms of the disease creates an opportunity to build model pipelines. These pipelines would be composed of mechanistic and stochastic within-host models of diseases, which outcomes can inform parameters for population-level models. This informed heterogeneity can alleviate the pressure of calibration or unknown parameter values in large models by leveraging known mechanisms of disease propagation in the body. When struggling with the unknown knowledge of disease progression and transmission, leaning on known biological mechanisms can provide insights to fill our gaps in knowledge.

\section*{Acknowledgments}
M.C.B. is is supported as a Fellow of the National Science Foundation under NRT award DGE-1735316.
L.H.-D. acknowledges financial support from the National Institutes of Health 2P20GM125498-06 Centers of Biomedical Research Excellence Award.

\bibliography{source}

\end{document}